# Two- vs. Three-Dose Optimization Under Sample Size Constraint

Linda Sun, Yixin Ren, Cong Chen*

Biostatistics and Research Decision Sciences, Merck & Co., Inc., Rahway, NJ 07065, USA
Corresponding author at: MAILSTOP UG-1CD44, 351 North Sumneytown Pike, North Wales, PA 19454, USA

E-mail address: cong_chen@merck.com

**Abstract**

Dose optimization is a hallmark of FDA's Project Optimus for oncology drug development. The number of doses to include in a dose optimization study depends on the totality of evidence, which is often unclear in early-phase development. With equal sample sizes per dose, carrying three doses is clearly more advantageous than two for optimization. In this paper, we show that, even when the total sample size is fixed, it is still preferable to carry three unless there is very strong evidence that one can be dropped. A mathematical approximation is applied to guide the investigation, followed by a simulation study to complement the theoretical findings. Semi-quantitative guidance is provided for practitioners, addressing both randomized and non-randomized dose optimization while considering population homogeneity.

Traditional cytotoxic chemotherapies typically exhibit a steep dose-response relationship, with the selected dose for Phase 2/3 development often at or near the maximum tolerable dose (MTD). This approach to dose selection is suboptimal for novel therapies with different or unclear dose-response relationships. To address this issue, the U.S. Food and Drug Administration (FDA) has launched Project Optimus, aiming to reform the dose optimization and selection paradigm with an emphasis of dose optimization in early phase development [1]. Unlike non-oncology areas, where a range of dose levels and regimens are routinely selected for optimization, achieving two-dose optimization after Optimus already represents a major leap forward. Asking for more than two doses would be a significant stretch in practice.

While PK/PD data and early efficacy and safety signals collected during dose escalation may suggest a reasonable dose range, the small sample size in this phase provides limited information to guide dose selection for a subsequent dose-optimization study. When considering three candidate doses (low, middle, high), a common question arises: should all three be included, or should only two be selected? Given a fixed total sample size, a two-arm study allows 50% more patients per arm compared to a three-arm study. However, excluding a third dose risks missing the optimal one. This decision involves complex clinical, scientific, and regulatory considerations [2]. We address it from a statistical perspective by noting that excluding the middle dose affects study power differently than excluding the low or high dose—an issue explored in this paper.

1. **A mathematical approximation**

Let $(\alpha, \beta)$ be the Type I/II errors of a two-arm study with equal allocation ratio for detecting an effect size of $\Delta$ for an early efficacy (or safety) endpoint. The overall sample size $N$ is approximately $4\frac{(F(\alpha)+F(\beta))^2}{\Delta^2}$ where $F(x) = \Phi^{-1}(x)$ is the standard normal quantile at $x$. For two studies targeting different effect sizes $(\Delta_1, \Delta_2)$ with different sample sizes $(N_1, N_2)$ at the same $\alpha$ level, the corresponding Type II errors $(\beta_1, \beta_2)$ satisfy the following relationship:

$$\frac{F(\alpha) + F(\beta_1)}{\Delta_1 \sqrt{N_1}} = \frac{F(\alpha) + F(\beta_2)}{\Delta_2 \sqrt{N_2}}$$

Let $r = \frac{\Delta_1 \sqrt{N_1}}{\Delta_2 \sqrt{N_2}}$, representing the relative signal strength adjusted for sample size, and assume it is greater than one without loss of generality. We show that $\frac{1-\beta_1}{1-\beta_2} \approx r$ when the study power $(1 - \beta)$ is in the range of a typical dose optimization study that is under powered at the usual alpha-level.

Rewrite the above equation as

$$F(\alpha) = \frac{F(\beta_1) - rF(\beta_2)}{r - 1}$$

Let $x = 1 - \beta_1$. Noting that $\beta_2 = 1 - x/r$ under the approximate power relationship, we define the righthand side function of the above equation as

$$G(x|r) = \frac{F(1 - x) - rF(1 - x/r)}{r - 1}$$

As shown in Figure 1, when both $x$ and $r$ fall within a range of practical interest, the value of $\Phi(G(x|r))$ lies approximately between 0.05 and 0.10. This implies that the corresponding powers to the two studies are proportional to $r$ when the dose comparison is conducted at the usual alpha-level for a Phase 2 study.

The seemingly surprising linear relationship arises directly from the fact that the normal CDF $\Phi(x)$ around zero can be approximated by a linear function with slope $\frac{1}{\sqrt{2\pi}}$ (the density function at zero). As a matter of fact, when power is between 25% and 75%, the difference between the linear function and $\Phi$ is less than 2% (Figure 2). Notice that the power function can be written as $1 - \beta = \Phi\left(\frac{\sqrt{N}}{2}\Delta + F(\alpha)\right)$, which can therefore be approximated by $\frac{1}{2\sqrt{2\pi}}\sqrt{N}\Delta + \frac{1}{\sqrt{2\pi}}F(\alpha) + 0.5$. When $\frac{1}{\sqrt{2\pi}}F(\alpha) + 0.5 \approx 0$, or when $\alpha \approx 0.1$, the power function would be approximately proportional to $\sqrt{N}\Delta$, just as implied in Figure 1. Because this paper aims to provide high-level guidance that can be used as a rule-of-thumb for practitioners, more precise analyses—while potentially of mathematical interest—are beyond its scope.

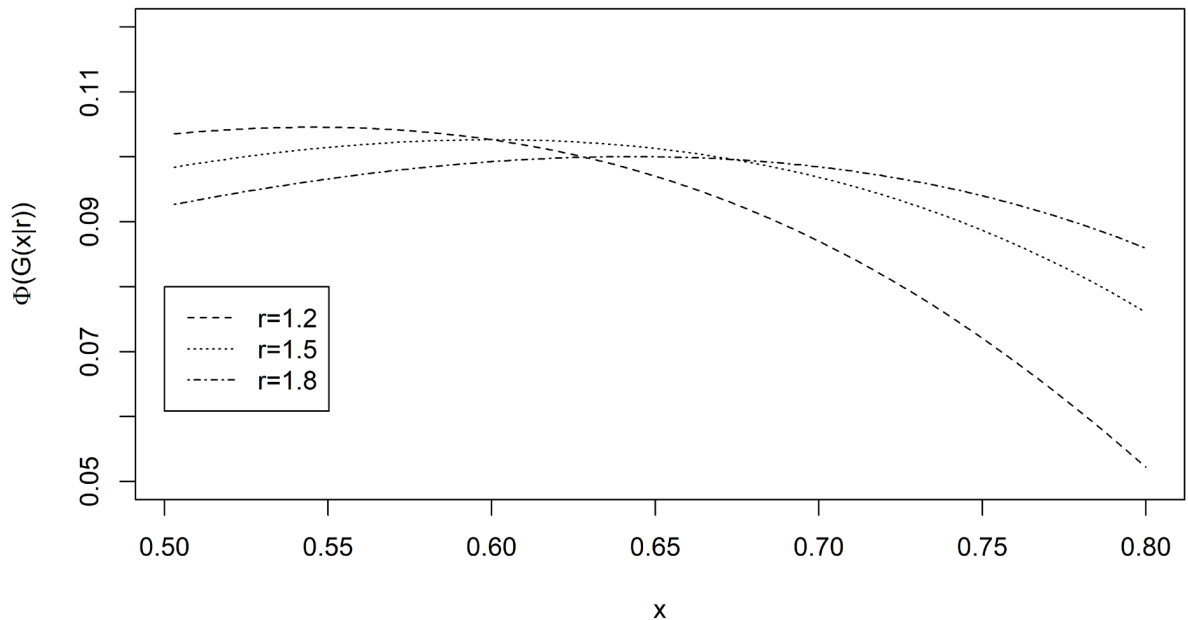

**Fig. 1.** Illustration of $\Phi(G(x|r))$ as a function of $x$ (power) for selected $r$, both in the range of practical interest. The fact that $\Phi(G(x|r))$ approximately falls between 0.05 and 0.10 suggests that power is approximately proportional to $r$ for underpowered studies.

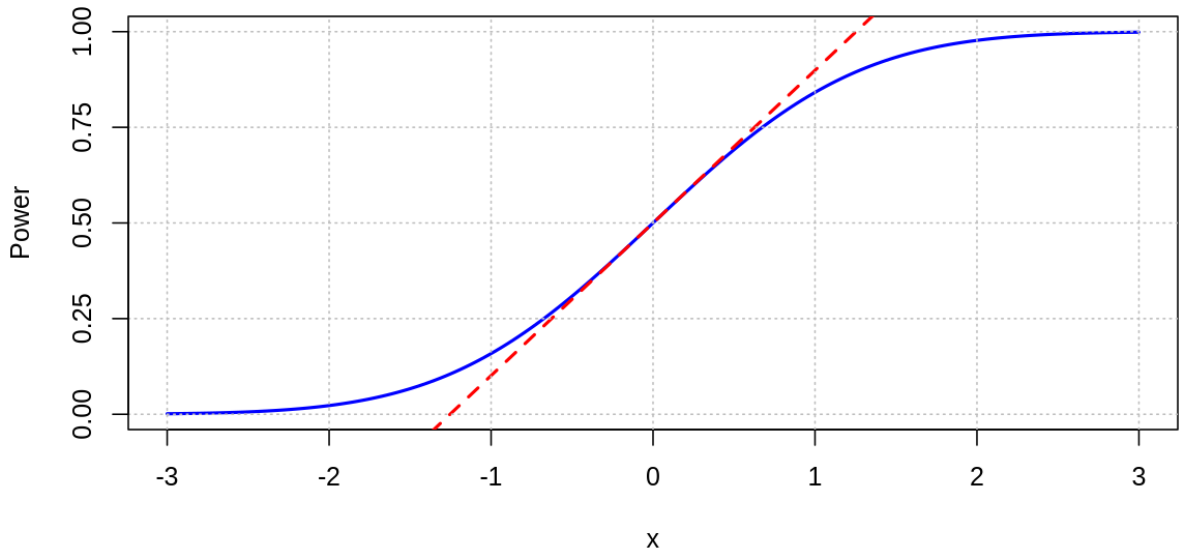

**Fig 2** CDF of standard normal distribution (in solid blue) and its linear approximation (in dotted red) at the vicinity of zero (i.e., when the underlying power is around 50%).

## 2. Practical implications

We now apply the mathematical approximation to dose optimization when there are three candidate doses for consideration (low, medium, and high) and the dose response relationship is assumed linear, which is the default and most natural assumption for power assessment in a dose optimization study.

Suppose a three-arm study has power $C$ for selecting a dose, primarily driven by the contrast between the high and low doses, as reflected in the Cochran-Armitage trend test. Based on the above discussion, if a two-arm study with the same total sample size is conducted, the power would be approximately $\sqrt{3/2}C$ when the low dose and high doses are included, and approximately $\sqrt{3/8}C$ when two adjacent doses (low and middle, or middle and high) are included. There is a prior belief that the middle dose is not optimal (e.g., comparable safety but inferior efficacy to high dose). Let this level of prior belief be denoted by $\lambda$. How large must $\lambda$ be to justify only including the low and high doses in the two-arm study? Ignoring $C$ on both sides of the inequality function, this question is equivalent to finding $\lambda$ such that:

$$\sqrt{3/2}\lambda + \sqrt{3/8}(1-\lambda) > 1 \text{ or } \lambda > \sqrt{8/3} - 1 \approx 0.63$$

Therefore, the middle dose may be excluded only when there is a reasonable level of prior belief (~60%) that it is not optimal.

We now consider a different question. Suppose a two-arm optimization study has power $C$ for detecting the treatment difference between two adjacent doses (low and middle, or middle and high). There is a prior belief that the optimal dose is between the two. Let this level of prior belief be denoted by $\lambda$. How large must $\lambda$ be to justify the two-arm study? Recognizing that including a dose unlikely to be optimal wastes resources, we can estimate $\lambda$ by solving the following inequality, using a similar approximation as described above:

$$\sqrt{2/3}\lambda + \sqrt{8/3}(1-\lambda) < 1 \text{ or } \lambda > 2 - \sqrt{3/2} \approx 0.78$$

Therefore, a stronger prior (~80%) is required to justify the study of two adjacent doses. Intuitively, although each arm has a smaller sample size, including the third dose enables a direct comparison between the low and high doses, which is statistically more powerful than comparing two adjacent ones.

## 3. A Simulation study

A simulation study is conducted to supplement the key message from the above mathematical evaluation. We implemented a Bayesian model-selection framework spanning four plausible dose-response shapes for efficacy (Figure 3), using priors calibrated to reflect realistic clinical expectations. We compared (i) a 3-dose design with N=30 per dose (L=low, M=middle, H=high) versus (ii) 2-dose comparisons (L vs H, M vs H or L vs M), holding total sample size constant. The primary metric was probability of correctly selecting the right dose (PCS) under each true shape.

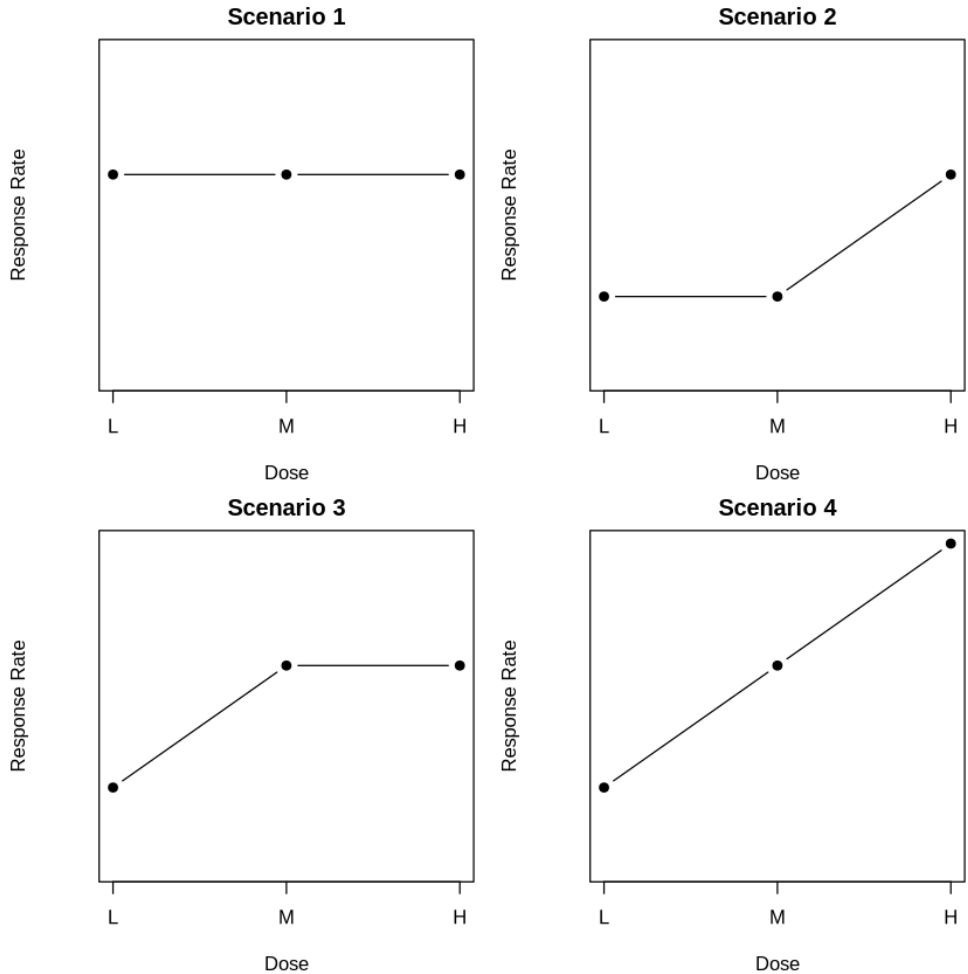

**Fig. 3** Four plausible dose-response shapes for the simulation study

Prior response rates $p_1, p_2, p_3$ for L, M, H doses are assumed to be: $p_1 \sim Beta(0.1, 1.9)$, $p_2 \sim Beta(0.6, 1.4)$, $p_3 \sim Beta(0.8, 1.2)$. The prior probabilities of the dose-response shapes are:

- If $p_2 - p_1 \leq 2.5\%$ and $p_3 - p_2 \leq 2.5\%$, count it as Scenario 1, Prior probability ~ 0%
- If $p_2 - p_1 \leq 2.5\%$ and $p_3 - p_2 > 2.5\%$, count it as Scenario 2, Prior probability ~ 20%
- If $p_2 - p_1 > 2.5\%$ and $p_3 - p_2 \leq 2.5\%$, count it as Scenario 3, Prior probability ~ 40%
- If $p_2 - p_1 > 2.5\%$ and $p_3 - p_2 > 2.5\%$, count it as Scenario 4, Prior probability ~ 40%

The posterior probabilities are calculated for 3-dose design for shape selection (see Appendix for details). The optimal dose selection rule is simple: choose the dose with higher efficacy; if efficacy is comparable, select the lower dose for safety consideration.

Figure 4 displays the PCS when the true dose-response is linear. The 3-dose design selected H with PCS ≈ 74%, versus 51% for the 2-dose comparison. The probability of selecting H is low in a 2-dose design because there are chances that the 2-dose design doesn't even include H dose (i.e., L and M are chosen to conduct the 2-dose optimization). If a 2-dose study consistently selects M and H as the randomized doses, the PCS may reach 78%; however, this comes at the cost of not being able to characterize the dose–response relationship. As expected, under the plateau scenario (results not shown here), the 2-dose and 3-dose designs yield comparable PCS for selecting M (the optimal dose). However, again, the 2-dose design cannot establish the dose–response relationship in a single study, whereas the 3-dose design can.

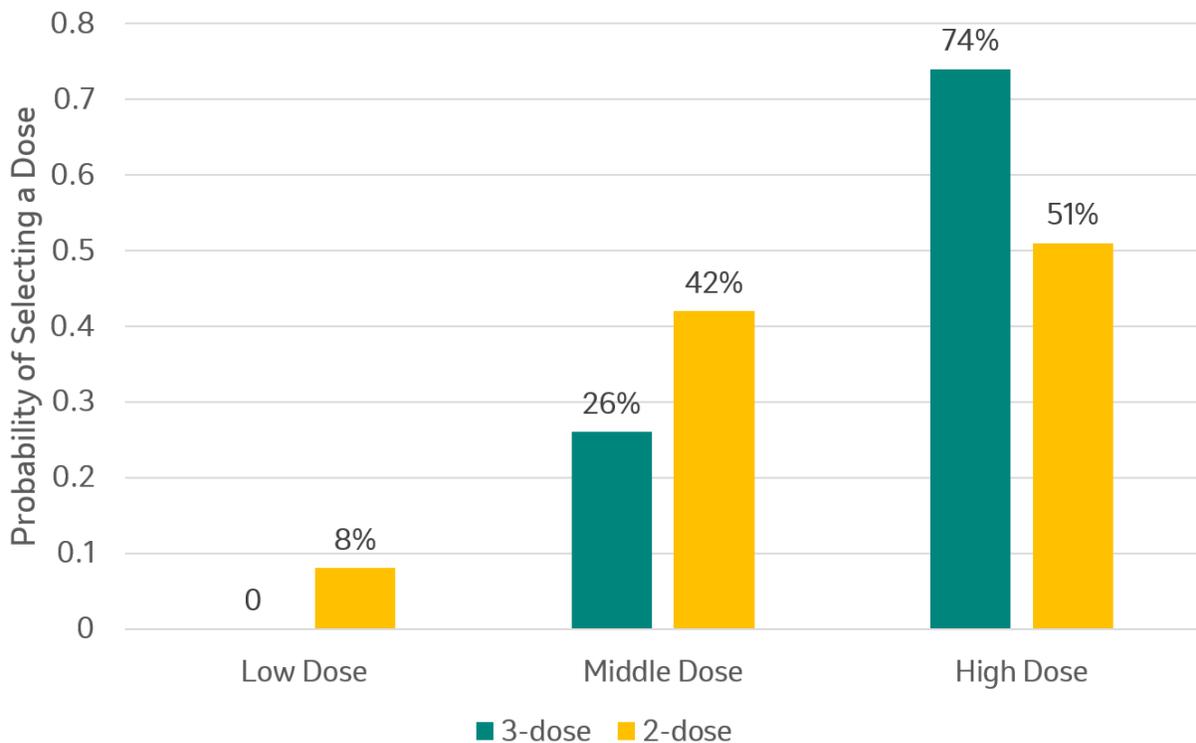

**Fig. 4** Probability of selecting a dose under linear assumption (Scenario 4)

4. **Practical guidance**

As shown in the previous two sections, unless there is strong confidence, the benefit of including three doses for optimization generally outweighs the drawback of having a smaller sample size per dose compared to two-dose optimization. Practically, 3-dose optimization can be achieved by backfilling in relatively homogeneous populations, dose randomization or a hybrid of the two. One example of the hybrid approach is to enroll additional patients at the low dose while randomizing only the middle and high doses when low-dose efficacy remains uncertain after escalation. This strategy offers more flexibility to discontinue the low dose for futility compared with randomizing all three doses.

We provide a 5-star rating system (Table 1) to compare various dose optimization strategies. The 3-dose optimization rates one star higher than 2-dose. Dose randomization rates two stars higher than backfilling which is not fully randomized, because backfilling may not show clear dose-response relationships due to baseline imbalances. However, if the dose escalation part of the dose-finding study is conducted in a homogeneous population, backfilling can save time and cost compared to dose randomization by leveraging patients already enrolled in dose escalation. The homogeneous population refers to a specific disease setting or a population with high expression of certain biomarkers. Table 1 shows that backfilling with a small number of patients in low-similarity populations is insufficient for dose optimization, and dose randomization is desired in this scenario. On the other hand, extended backfilling for $\geq 2$ doses with larger sample size in a high-similarity (i.e., homogeneous) population could provide comparable dose optimization information to 2-dose randomization. Other practical considerations include the number of indications and regions for dose optimization. If two indications or regions are desired, a viable strategy is to use backfilling for one and dose randomization for the other.

Table 1. A Practical 5-star Rating System for Phase 3 Dose Readiness

|  |  | Dose randomization | | |
|---|---|---|---|---|
|  |  | No | 2-dose (30-40/dose) | 3-dose (20-30/dose) |
| Expansion of dose escalation | No | - | ** | *** |
| | Backfill (low similarity, 10/dose) | * | *** | **** |
| | Extended backfill (incl. backfill) (moderate similarity, 20+/dose, ≥2 doses) | ** | **** | ***** |
| | Extended backfill (incl. backfill) (high similarity, 20+/dose, ≥2 doses) | *** | ***** | ******⁺ |

## 5. Discussions

The key finding in this paper align with the intuition of experienced drug developers: when selecting only two doses for optimization, it is generally more informative to space them apart. However, selecting two adjacent doses requires stronger confidence on excluding the third dose. When confidence is lower, a three-arm study may be considered, as it not only facilitates dose selection but also characterizes the dose-response curve. Mathematical approximation is used to derive the findings under a linear dose-response relationship. A simulation study based on a Bayesian framework is conducted to supplement the key message from the mathematical evaluation.

In practice, it is essential to incorporate the full spectrum of data generated from dose escalation and backfilling [3–4] into the dose selection process. These complementary data sources help refine dose selection by narrowing the range of plausible doses and guiding the choice of dose levels for optimization, particularly when aiming to select two with confidence. To balance statistical rigor with practical feasibility, semi-quantitative guidance is provided for practitioners, addressing both randomized and non-randomized dose optimization while considering population homogeneity.

A control arm may be included in a dose-randomization study as it can not only support dose optimization but also strengthen proof-of-concept, especially when preliminary data leave significant uncertainty about the drug activity. Moreover, this study can be seamlessly integrated into an adaptive Phase 2/3 design, allowing the data to contribute directly to the Phase 3 primary analysis and potentially improving the overall efficiency of the development program [5].

**References**

1. Fourie Zirkelbach J, Shah M, Vallejo J, Cheng J, et al. Improving dose-optimization processes used in oncology drug development to minimize toxicity and maximize benefit to patients. *J Clin Oncol*. 2022 Sep 12: JCO2200371.
2. FDA Guidance. Optimizing the Dosage of Human Prescription Drugs and Biological Products for the Treatment of Oncologic Diseases 2024. https://www.fda.gov/regulatory-information/search-fda-guidance-documents/optimizing-dosage-human-prescription-drugs-and-biological-products-treatment-oncologic-diseases?utm_source=chatgpt.com.
3. Zhao Y, Yuan Y, Korn EL, Freidlin B. Backfilling Patients in Phase I Dose-Escalation Trials Using Bayesian Optimal Interval Design (BOIN). *Clin Cancer Res*. 2024 Feb 16;30(4):673-679. doi: 10.1158/1078-0432.CCR-23-2585.
4. Wang J, Li Z, Yung G, Fridlyand J, Li C, and Zhu J. (2025). Evaluations of Backfill Strategies in Dose Optimization through Simulation Studies for Phase I Trials in Oncology. *Stat Biopharm Res* 2025, 1–11. https://doi.org/10.1080/19466315.2025.2529424.
5. Chen C, Huang M, Zhang X. Adaptive Phase 2/3 design with dose optimization. *Contemp. Clin. Trials*. 2025. To appear.

# Appendix

For 3-dose design, choose the shape with the highest posterior probability. Below is an example of posterior probability of shapes using n=30/dose.

**Table 1.** Posterior probability of four shapes using N=30 per dose

| Observed number of responders for L, M, H doses | Observed ORR for L, M, H doses | Posterior Probability | | | |
|---|---|---|---|---|---|
| | | S1 | S2 | S3 | S4 |
| S1: (6, 6, 6) | (20%, 20%, 20%) | 19% | 36% | 41% | 5% |
| S2: (3, 3, 6) | (10%, 10%, 20%) | 4% | 55% | 19% | 22% |
| S3: (3, 6, 6) | (10%, 20%, 20%) | 3% | 16% | 60% | 21% |
| S4: (3, 6, 9) | (10%, 20%, 30%) | 0.5% | 18% | 27% | 55% |

For 2-dose design,

    S1: choose L and M dose to conduct the dose randomization – 0%
    S2: choose L and H dose – 20%
    S3: choose L and M dose – 40%
    S4: choose M and H dose – 40%

Choose the higher dose if posterior probability $(p_h - p_l > 2.5\%) > 50\%$.